\documentclass{elsart1p}
\usepackage{graphics}
\usepackage{graphicx}
\usepackage{epsfig}
\usepackage{amssymb}

\begin{document}

\begin{frontmatter}

\title{Symmetries and Symmetry Breaking Patterns in QCD:\\
Chiral and Deconfinement Transitions}
\author{Wolfram Weise}\footnote{Supported in part by BMBF, GSI and by the DFG cluster of excellence Origin and Structure of the Universe.}
\address{Physik-Department, Technische Universit\"at M\"unchen, D-85747 Garching, Germany}

\begin{abstract}
The symmetry breaking pattern of QCD features two seemingly disconnected
phenomena: the spontaneous breakdown of the $Z(3)$ center symmetry 
in the deconfinement transition of pure-gauge QCD, and the spontaneous breaking of
chiral $SU(N_f)\times SU(N_f)$ symmetry in the limit of  $N_f$ massless quark flavours.
The dynamical entanglement of these symmetries is displayed in the framework of a 
schematic model (the PNJL model) in comparison with results from Lattice QCD. Extensions to non-zero baryon chemical potential are discussed.
\end{abstract}
\end{frontmatter}

\section{Introduction}
\label{Intro}

Confinement and spontaneous chiral symmetry breaking in QCD are governed by  two basic symmetry principles:

{\cal i)} The symmetry associated with the center $Z(3)$ of the local $SU(3)_c$ color gauge group is exact in the limit of pure gauge QCD, realized for {\it infinitely heavy} quarks. In the high-temperature, deconfinement phase of QCD this $Z(3)$ symmetry is spontaneously broken, with the Polyakov loop acting as the order parameter.

{\cal ii)} Chiral $SU(N_f)_R\times SU(N_f)_L$ symmetry is an exact global symmetry of QCD with $N_f$ {\it massless} quark flavors.  In the low-temperature (hadronic) phase this symmetry is spontaneously broken down to the flavor group $SU(N_f)_V$ (the isospin group for $N_f = 2$ and the ``eightfold way" for $N_f = 3$). As a consequence there exist $2N_f + 1$ pseudoscalar Nambu-Goldstone bosons and the QCD vacuum hosts a quark condensate. 

Confinement implies spontaneous chiral symmetry breaking, whereas the reverse is not necessarily true. Whether and under which conditions the chiral and deconfinement transitions coincide, as it appears to be the case in recent lattice QCD computations with almost physical quark masses,  is a fundamental issue. 

\section{Chiral and deconfinement transitions}

{\it Chiral condensate and Polyakov loop}. The order parameter of spontaneously broken chiral symmetry is the quark condensate, $\langle \bar{q} q \rangle$.  The  disappearence of this condensate, by its melting above a transition temperature $T_c$, signals the restoration of chiral symmetry in Wigner-Weyl realization. The transition from confinement to deconfinement in QCD is likewise controlled by an order parameter, the Polyakov loop. A non-vanishing Polyakov loop $\Phi$ reflects the spontaneously broken $Z(3)$ symmetry characteristic of the deconfinement phase. The Polyakov loop vanishes in the low-temperature, confinement sector of QCD.

%figure-----------------------------------------------------------------------------------------------------------------------------
\begin{figure}[htb]
\begin{minipage}[t]{7cm}
\includegraphics[width=5.9cm]{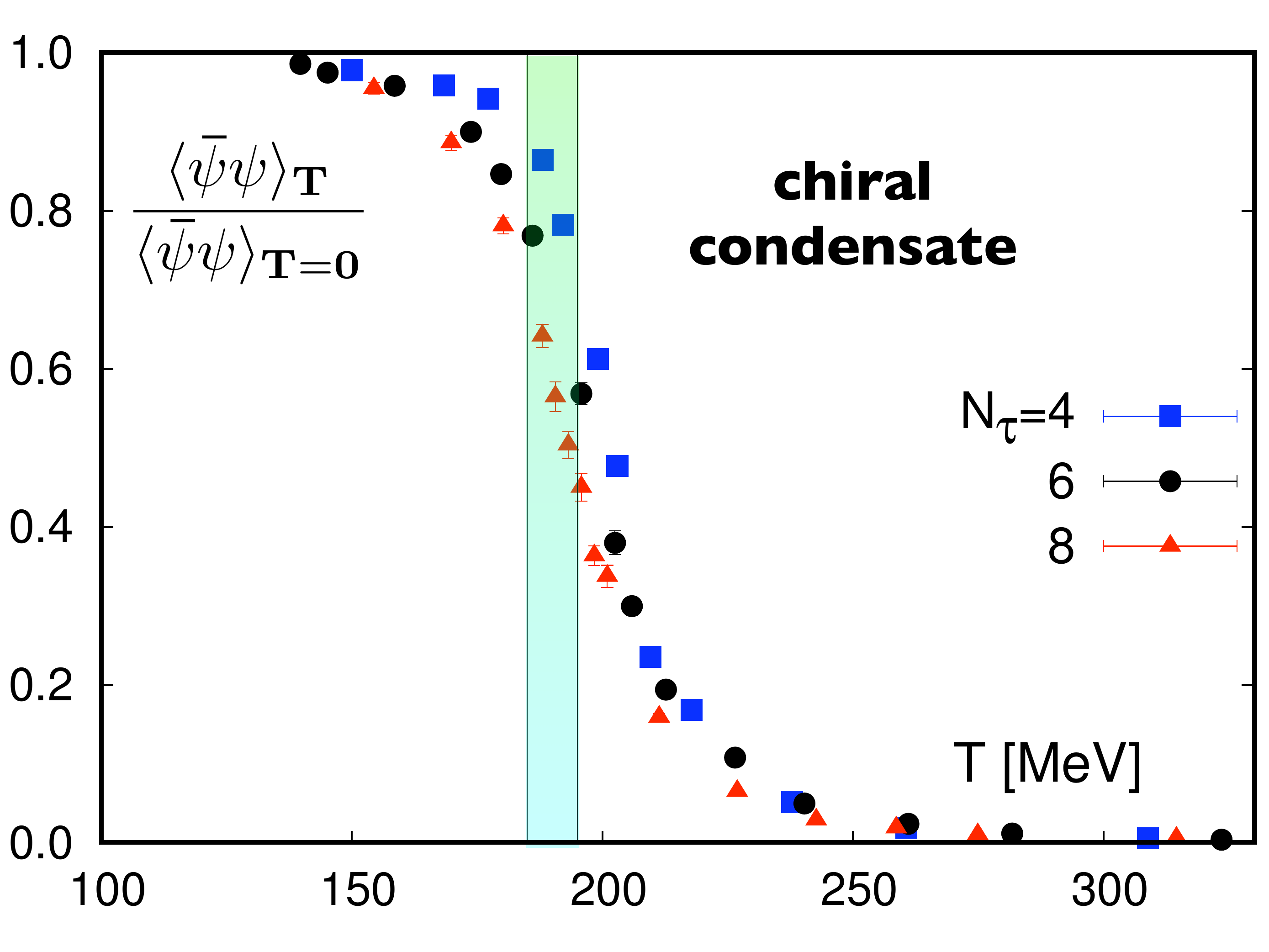}
\label{fig:1}
\end{minipage}
\hspace{\fill}
\begin{minipage}[t]{7cm}
\includegraphics[width=5.9cm]{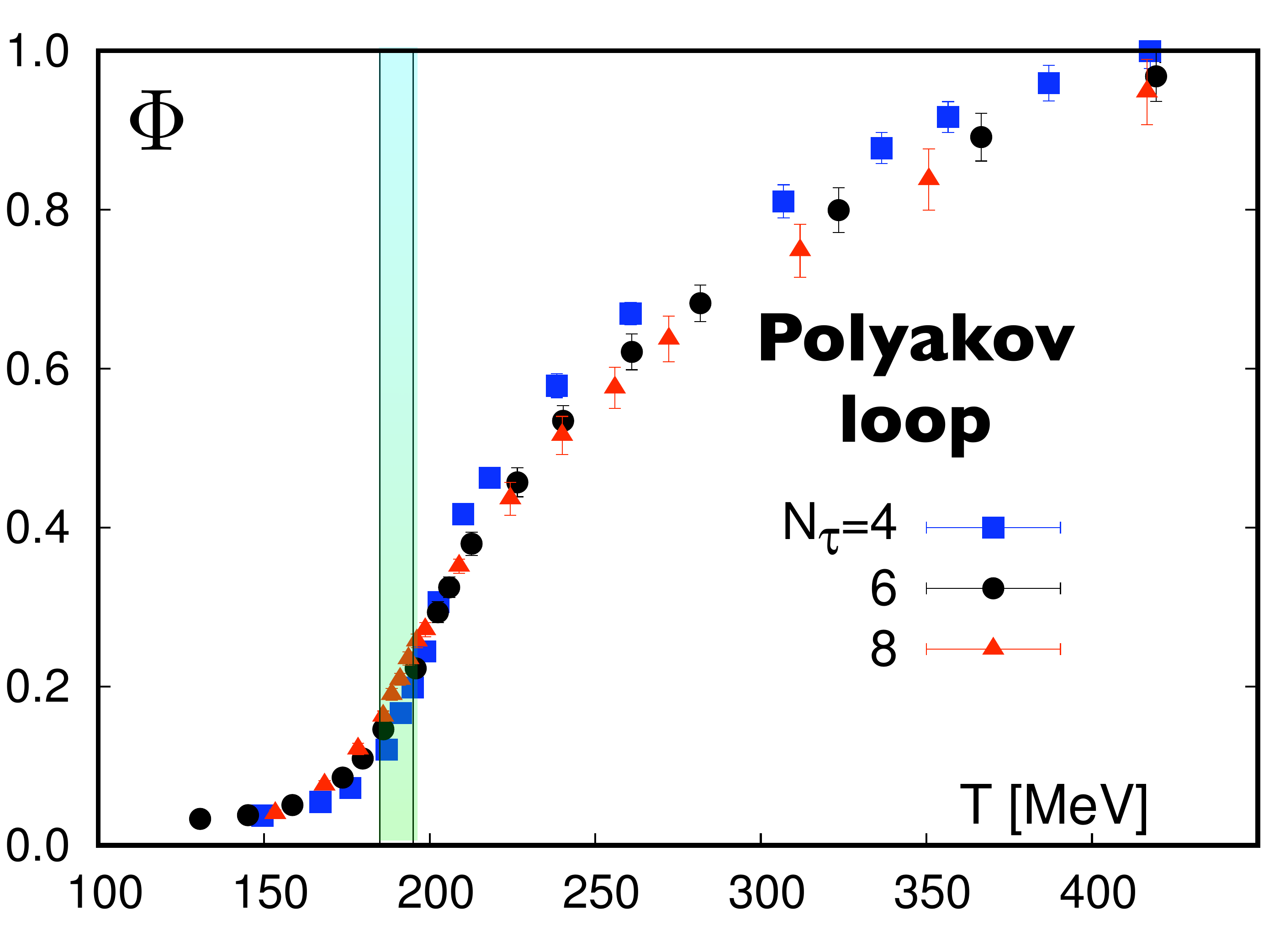}
\end{minipage}
\caption{Lattice QCD results ($N_f = 2+1$) of chiral condensate (left) und Polyakov loop (right) as functions of temperature \cite{Ch2008}. Different data sets correspond to different number $N_\tau$ of lattice points along the Euclidean time axis.} 
%\hspace{\fill}
\end{figure}
%
%figure-----------------------------------------------------------------------------------------------------------------------------

There is no principal reason why the deconfinement and chiral transitions should occur at the same temperature. Nonetheless, recent results of lattice QCD thermodynamics \cite{Ch2008} with 2+1 flavors (at zero baryon chemical potential) indicate just that (see Fig.1), with a common transition temperature $T_c \simeq 190$ MeV. A deeper understanding of this observation is of fundamental importance, also in view of the fact
that earlier lattice simulations \cite{Aoki2006} still found a displacement between chiral and deconfinement temperatures (although a more detailed assessment of systematic uncertainties might resolve this apparent contradiction).

One notes that the chiral and deconfinement transitions as shown in Fig.1 are not phase transitions but smooth crossovers, so there is no critical temperature in the strict sense. It is nevertheless possible to define a transition temperature band around the maximum slope of either the condensate $\langle\bar{q}q\rangle_T$ or the Polyakov loop $\Phi(T)$. 

Two limiting cases are of interest in this context. In pure gauge QCD, corresponding to infinitely heavy quarks, the deconfinement transition is established - at least in lattice QCD - as a first order phase transition with a critical temperature of about $270$ MeV. In the limit of massless $u$ and $d$ quarks, on the other hand, the isolated chiral transition appears as a second order phase transition at a significantly lower critical temperature. This statement is based on calculations using Nambu - Jona-Lasinio (NJL) type models which incorporate the correct spontaneous chiral symmetry breaking mechanism but ignore confinement. The step from first or second order phase transitions to crossovers is understood as a consequence of explicit symmetry breaking. The $Z(3)$ symmetry is explicitly broken by the mere presence of quarks with non-infinite masses. Chiral symmetry is explicitly broken by non-zero quark masses. But the challenging question remains how the chiral and deconfinement transitions get dynamically entangled in just such a way that they finally occur within a common transition temperature interval.

{\it The PNJL model}. Insights concerning this issue can be gained from a model based on a minimal synthesis of the NJL-type spontaneous chiral symmetry breaking mechanism and confinement implemented through Polyakov loop dynamics. This PNJL model \cite{Fu2003,RTW2006} is specified by the following action:
\begin{eqnarray}
{\cal S} = \int_0^{\beta=1/T}d\tau \int_V d^3x\left[\psi^\dagger\partial_\tau\psi-{\cal H}(\psi,\psi^\dagger,\phi)\right] - {V\over T}\,{\cal U}(\Phi,T)~.
\end{eqnarray}
It introduces the Polyakov loop, $\Phi = N_c^{-1}\,Tr\exp(i\phi/T),$ with a homogeneous temporal gauge field, $\phi = \phi_3\lambda_3 + \phi_8\lambda_8\in SU(3)$, coupled to the quarks. The dynamics of $\Phi$ is controlled by a $Z(3)$ symmetric effective potential ${\cal U}$, designed such that it reproduces the equation of state of pure gauge lattice QCD with its first order phase transition at a critical temperature of 270 MeV. The field $\phi$ acts as a potential on the quarks represented by the flavor doublet (for $N_f = 2$) or triplet (for $N_f = 3$) fermion field $\psi$. The Hamiltonian density in the quark sector is ${\cal H} = -i\psi^\dagger(\vec{\alpha}\cdot\vec{\nabla} +\gamma_4\,\hat{m} - \phi)\psi + {\cal V}(\psi,\psi^\dagger)$, with the quark mass matrix $\hat{m}$ and a chiral $SU(N_f)_L\times SU(N_f)_R$ symmetric interaction ${\cal V}$. 

Earlier two-flavor versions of the PNJL model \cite{Fu2003,RTW2006,RRW2007} have still used a local four-point interaction of the classic NJL type, requiring a momentum space cutoff to regularize loops. A more recent version \cite{HRCW2009} using a non-local interaction does not require an artificial cutoff. It generates instead a momentum dependent dynamical quark mass, $M(p)$, along with the non-vanishing quark condensate. A further extension to three quark flavors includes a $U(1)_A$ breaking term implementing the axial anomaly of QCD. 

With the input fixed at zero temperature by well-known properties of the pseudoscalar mesons, the thermodynamics of the PNJL model can then be investigated with focus on the symmetry breaking pattern and on the intertwining of chiral dynamics with that of the Polyakov loop. The primary role of the Polyakov loop and its coupling to the quarks is to supress the thermal distribution functions of color non-singlets, i.e. quarks and diquarks, as the transition temperature $T_c$ is approached from above. Color singlets, on the other hand,  are left to survive below $T_c$.
%figure-----------------------------------------------------------------------------------------------------------------------------
\begin{figure}[t]
\begin{minipage}[t]{7cm}
\includegraphics[width=6cm]{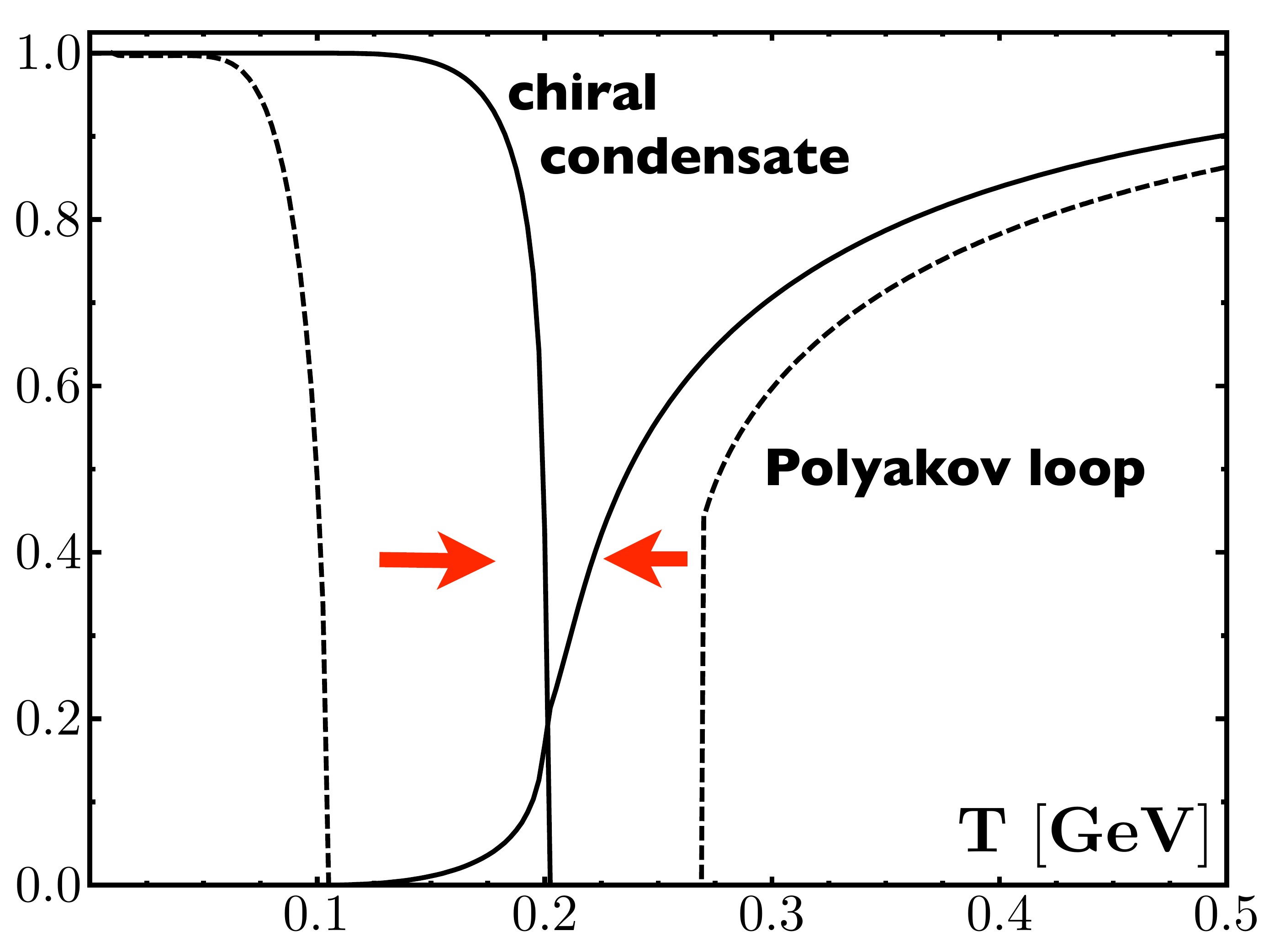}
\label{fig:2a}
\end{minipage}
\hspace{\fill}
\begin{minipage}[t]{7cm}
\includegraphics[width=6cm]{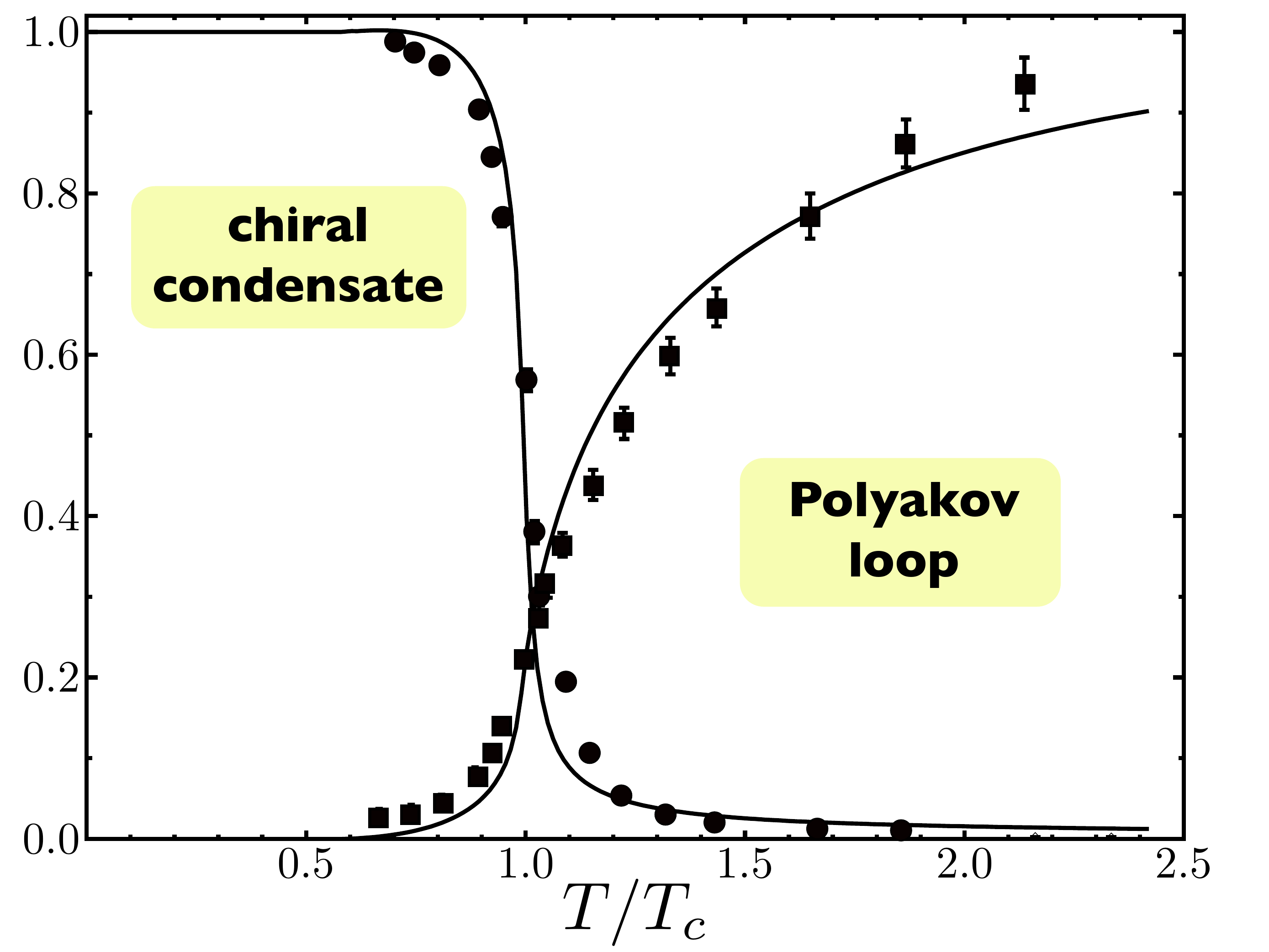}
\end{minipage}
\caption{PNJL model calculations of chiral and deconfinement transitions \cite{RRW2007,HRCW2009}. See text for  explanations.}
\label{fig:4}
%\hspace{\fill}
\end{figure}
%
%figure-----------------------------------------------------------------------------------------------------------------------------

A remarkable dynamical entanglement of the chiral and deconfinement transitions is then observed, as demonstrated in Fig.\ref{fig:4} (left) for the two-flavor case. In the absence of the Polyakov loop the quark condensate (left dashed line), taken in the chiral limit, shows the expected 2nd order chiral phase transition, but at a temperature way below and far separated from the 1st order deconfinement transition controlled by the pure-gauge Polyakov loop effective potential ${\cal U}$ (right dashed line). Once the coupling of the Polyakov loop field to the quark density is turned on, the two transitions move together and end up at a common transition temperature around 0.2 GeV. The deconfinement transition becomes a crossover (with $Z(3)$ symmetry explicitly broken by the coupling to the quarks), while the chiral phase transition remains 2nd order until non-zero $u$ and $d$ quark masses, $m_{u,d}  \simeq 4$ MeV,  induce a crossover transition as well.  

The right side of Fig.\ref{fig:4} shows the two-flavor PNJL result \cite{HRCW2009} together with $N_f=2+1$ lattice data \cite{Ch2008}. A direct comparison is clearly not appropriate but the similarity of the crossover transition patterns is striking, given the simplicity of the model and the fact that these results are derived from a mean-field approach. Further important steps toward systematic investigations beyond mean field are presently being pursued.

\section{Scenarios at finite baryon density}

{\it The quest for the critical point}. Undoubtedly a prime challenge in the physics of strong interactions is the exploration of the QCD phase diagram at non-zero baryon density, extending from normal nuclear matter all the way up to very large quark chemical potentials $\mu$ at which color superconducting phases are exŸected to occur. PNJL calculations at finite $\mu$ give a pattern of the chiral order parameter in the $(T,\mu)$ plane showing the onset of a first order transition with a critical point and a first-order transition line extending down to a quark chemical potential $\mu \sim$ 0.3 - 0.4 GeV at $T = 0$.

Several important questions are being raised in this context. The first one concerns the existence and location of the critical point. Extrapolations from lattice QCD, either by Taylor expansions around $\mu =0$
\cite{FK2006} or by analytic continuation from imaginary chemical potential \cite{FP2008}, have so far not reached a consistent conclusion. A second question relates to the sensitivity  of the first order transition line in the phase diagram with respect to the axial $U(1)_A$ anomaly in QCD. This issue has been addressed in Ref.\cite{YHB2007} and it was pointed out that, depending on details of the axial $U(1)_A$ breaking interaction, a second critical point might appear such that the low-temperature evolution to high density is again just a smooth crossover, or the first order transition might disappear altogether and give way to a smooth crossover throughout.

\begin{figure}[htb]
\begin{minipage}[t]{5.5cm}
\includegraphics[width=5.5cm]{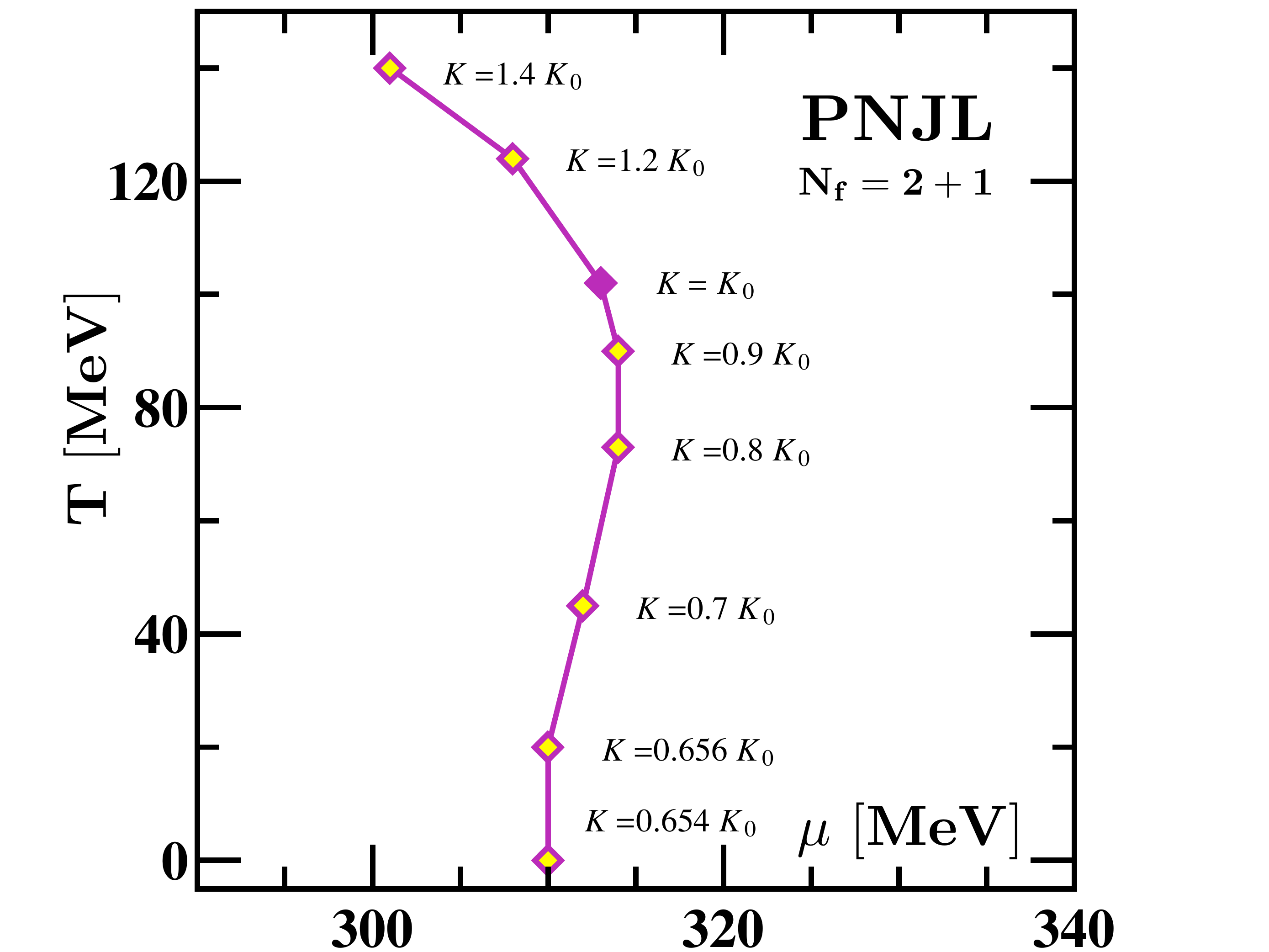}
\caption{Three-flavour PNJL calculation \cite{BHRW2009} showing the dependence of the critical point on the strength $K$ of the axial $U(1)$ breaking interaction; $K=K_0$ is the point at which the $\eta'$ mass in vacuum is reproduced.} 
\label{fig:5}
\end{minipage}
\hspace{\fill}
\begin{minipage}[t]{7cm}
\includegraphics[width=6cm]{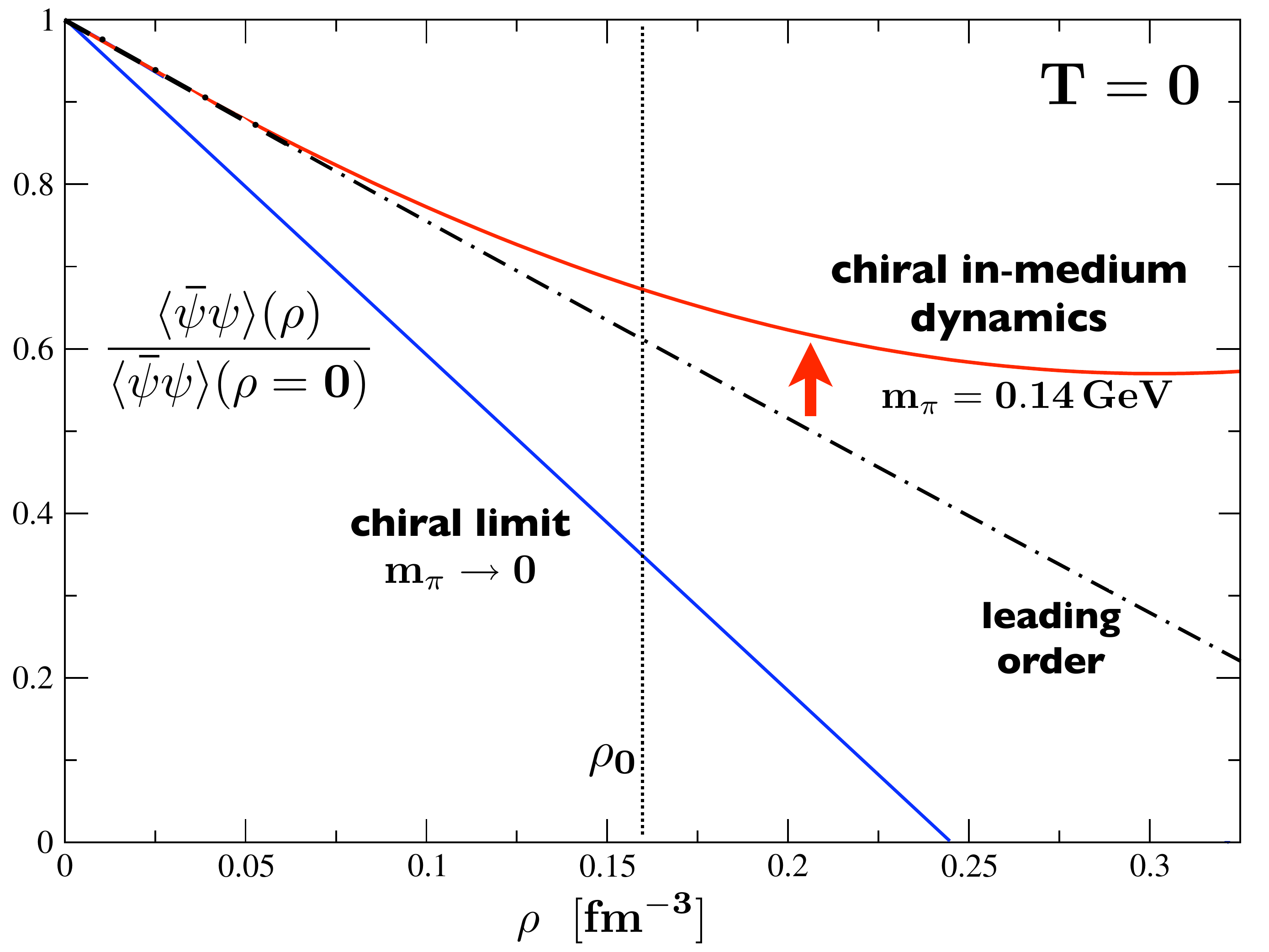}
\caption{Density dependence of the chiral condensate in symmetric nuclear matter \cite{KHW2008}. Dot-dashed curve: leading order term using $\sigma_N = 50$ MeV. Upper curve: full in-medium chiral dynamics result at three-loop order. Lower curve: chiral limit with vanishing pion mass.} 
\label{fig:6}
\end{minipage}
\end{figure}

An impression of the explicit dependence of the critical point on the axial anomaly can be obtained using the three-flavor PNJL model with inclusion of a $U(1)_A$ breaking Kobayashi-Maskawa-`t Hooft determinant interaction and varying the coupling strength $K$ of this interaction \cite{Fu2008,BHRW2009}. It turns out that the location of the critical point in the phase diagram varies indeed strongly with $K$ and may even disappear altogether below a certain value of $K$ (see Fig.\ref{fig:5}). A closely related question is how the mass of the $\eta\,'$ meson behaves in a dense baryonic environment. 
%figure-----------------------------------------------------------------------------------------------------------------------------
%figure-----------------------------------------------------------------------------------------------------------------------------

From the variety of existing model calculations (including those using the PNJL model) one might draw the presumably premature conclusion that critical phenomena occur already at a density scale not much higher than that of normal nuclear matter. However, all these models are not capable of working with the proper degrees of freedom around and below a baryon chemical potential $\sim$ 1 GeV (corresponding to quark chemical potentials around 0.3 GeV). Approaching this density scale from below, it is obvious that constraints from what we know about the nuclear matter equation of state must be seriously considered.

{\it Constraints from nuclear matter}. Chiral effective field theory is not only the low-energy realization of QCD in the meson and single-baryon sectors. It is also a basis for dealing with the nuclear many-body problem in terms of in-medium chiral perturbation theory \cite{FKW2005}. In this approach, chiral one- und two-pion exchange processes in the nuclear medium are treated explicitly while unresolved
short-distance dynamics is encoded in contact terms. Present calculations are performed up to three-loop order in the energy density. Three-body interactions emerge and play a significant role in this framework. The pion mass $m_\pi$, the nuclear Fermi momentum $p_F$ and the
mass splitting between nucleon and $\Delta(1232)$ are all comparable scales. Therefore the relevant,
active degrees of freedom are pions, nucleons and $\Delta$ isobars. Intermediate range two-pion exchange interactions produce van der Waals - like forces involving the large spin-isospin polarizablity
of the individual nucleons, and the Pauli principle acts on intermediate nucleon states in two-pion exchange processes. 

This scenario leads to a realistic nuclear matter equation of state \cite{FKW2005} with a liquid-gas first order phase transition and a critical temperature of about 15 MeV, close to the range of empirical values extracted for this quantity. This is so far the only well established part of the phase diagram of strongly interacting matter at finite density and low temperature. The truncation in the chiral expansion of the pressure in powers of the Fermi momentum, $p_F$, implies that these calculations can be trusted up to about twice the density of normal nuclear matter, i.e. for $p_F \lesssim 0.3$ GeV $<< 4\pi f_\pi$ with $f_\pi = 0.09$ GeV the pion decay constant. 

In a nuclear equation of state based on chiral dynamics the pion mass enters explicitly (or, equivalently, the quark mass according to the Gell-Mann - Oakes - Renner relation $m_\pi^2 f_\pi^2 = -m_q\langle\bar{\psi}\psi\rangle$). Equiped with such an equation of state one can now ask the following question: how does the chiral condensate extrapolate
to baryon densities exceeding those of normal nuclear matter? In-medium chiral effective field theory
gives the following answer:
\begin{eqnarray}
{\langle\bar{\psi}\psi\rangle_\rho\over\langle\bar{\psi}\psi\rangle_0} = 1&-&{\rho\over f_\pi^2}{\sigma_N\over m_\pi^2}\left(1 - {3\,p_F^2\over 10\,M_N^2} + \dots\right) + {\rho\over f_\pi^2}{\partial\over \partial m_\pi^2}\left({E_{int}(p_F)\over A}\right)~. 
\end{eqnarray}
The second term on the r.h.s., with its leading linear dependence on density, is the contribution from a free Fermi gas of nucleons, with the pion-nucleon sigma term $\sigma_N\simeq 0.05$ GeV and non-static corrections. The third term involves the pion mass dependence of the interaction energy per nucleon, $E_{int}/A$. This term features prominently the two-pion exchange interaction in the nuclear medium  including Pauli principle corrections, and also three-nucleon forces based on two-pion exchange.

The dashed-dotted curve in Fig.\ref{fig:6} shows the pronounced leading linear reduction in the magnitude of the chiral condensate with increasing density. This holds in the absence of correlations between the nucleons. Up to about the density of normal nuclear matter, this term dominates, whereas the interaction part tends to delay the tendency towards chiral restoration when the baryon density is further increased. This behaviour is sensitive to the actual value of the pion mass. In the chiral limit, $m_\pi\rightarrow 0$, with stronger attraction in the NN force at intermediate ranges, the trend is reversed and the rapidly dropping condensate would now lead to the restoration of chiral symmetry at relatively low density: nuclear physics would look completely different if the pion were an exactly massless Nambu-Goldstone boson. The influence of  explicit chiral symmetry breaking through the small but non-zero $u$ and $d$ quark masses on qualitative properties of nuclear matter is quite remarkable. 

These results demonstrate that known properties of a realistic nuclear equation of state must be considered as important constraints for extrapolations to higher densities, at least at low temperatures.
PNJL type models work with quarks as independent quasiparticles.  They do not account for those parts of the QCD phase diagram that are governed by color singlet baryons as relevant degrees of freedom. 
The recent very interesting discussion  \cite{McL2009} of a quarkyonic sector in the phase diagram at moderate quark chemical potentials should not miss these constraints.

\end{document}